\documentclass[aps,reprint,amsmath,amssymb,groupaddress,superscriptaddress,pra]{revtex4-2}

\usepackage{newtxtext}
\usepackage{newtxmath}
\usepackage{graphicx}
\usepackage[utf8]{inputenc}
\usepackage{amsmath}
\usepackage{subfigure}
\usepackage{amsfonts}
\usepackage{amssymb}
\usepackage{bm}
\usepackage{dcolumn}
\usepackage{color}
\usepackage[T1]{fontenc}
\usepackage{booktabs, array, float, tabularx, lipsum, multirow, mathtools}
\graphicspath{{figs/}{figsgaoerb/}} 
\usepackage{soul}
\usepackage{CJKutf8}
\usepackage{xcolor}
\usepackage{booktabs}

\usepackage[colorlinks,citecolor=blue,linkcolor=blue,anchorcolor=blue,urlcolor=blue]{hyperref}

\begin{document}

\begin{CJK*}{UTF8}{gbsn}
\title{Carnot Meets Quantum Information: A Thermal Machine Driven by Non-orthogonal State Discrimination}
\author{Tan-Ji Zhou (周谭吉)}
\altaffiliation{These authors contributed equally to this work.}
\affiliation{Graduate School of China Academy of Engineering Physics, Beijing, 100193, China}

\author{Yun-Qian Lin (林蕴芊)}
\altaffiliation{These authors contributed equally to this work.}
\affiliation{School of Physics and Astronomy, Beijing Normal University, Beijing, 100875, China}
\affiliation{Key Laboratory of Multiscale Spin Physics (Ministry of Education), Beijing Normal University, Beijing 100875, China}

\author{Yu-Han Ma (马宇翰)}
\email{yhma@bnu.edu.cn}
\affiliation{School of Physics and Astronomy, Beijing Normal University, Beijing, 100875, China}
\affiliation{Key Laboratory of Multiscale Spin Physics (Ministry of Education), Beijing Normal University, Beijing 100875, China}

\author{C. P. Sun (孙昌璞)}
\email{suncp@gscaep.ac.cn}
\affiliation{Graduate School of China Academy of Engineering Physics, Beijing, 100193, China}
\begin{abstract} 

While the impossibility of perfectly identifying non-orthogonal states is a cornerstone of quantum information science, they can nevertheless be distinguished with a finite error probability. Here, we propose a two-reservoir quantum machine driven by such non-orthogonal-state discrimination to map its functional boundaries across the parameter space of the state overlap $\mu$ and the Carnot efficiency $\eta_C$. Within this $\eta_C$-$\mu$ plane, the machine exhibits phase-transition-like functional switching among a pure heat-engine phase, a mixed phase, and a dissipative phase. We identify critical thresholds governing these transitions: strong thermal driving ($\eta_C>0.5$) unconditionally guarantees positive work extraction, whereas weak driving ($\eta_C \lesssim 0.13$) induces an anomalous reentrant transition, where increasing $\mu$ unexpectedly restores engine functionality after a purely dissipative regime. Our results explicitly demonstrate how quantum mechanics and thermodynamics jointly constrain information-to-energy conversion.
\end{abstract}

\maketitle
\textit{Introduction.---} While quantum mechanics strictly forbids the perfect discrimination of
non-orthogonal states, it nevertheless permits their imperfect discrimination with a finite error probability \citep{helstrom1969Quantumdetectionestimation, yuen1975Optimumtestingmultiple, helstrom1976quantum, bae2013Structureminimumerrorquantum, bae2015Quantumstatediscrimination}. Extracting such quantum information inextricably links information processing to thermodynamics \citep{szilard1929UberEntropieverminderungthermodynamischen, alicki2004ThermodynamicsQuantumInformation, sagawa2009MinimalEnergyCost, jacobs2009Secondlawthermodynamics, maruyama2009ColloquiumphysicsMaxwells, sagawa2010GeneralizedJarzynskiEquality, hilt2011Landauersprinciplequantum, esposito2011SecondlawLandauer, sagawa2012Fluctuationtheoreminformation, reeb2014ImprovedLandauerprinciple, parrondo2015Thermodynamicsinformation, alhambra2016FluctuatingWorkQuantum, naghiloo2018InformationGainLoss, ptaszynski2019ThermodynamicsQuantumInformation}. Landauer's principle explicitly quantifies this relationship by imposing a strict energetic cost on information erasure \citep{landauer1961Irreversibilityheatgeneration}. This thermodynamic bound universally governs information machines, such as the Szilard engine, which extract work by utilizing a Maxwell's demon to discriminate the state of a working substance \citep{bennett1982Thermodynamicscomputationreview, quan2006MaxwellsDemonAssisted, dong2011QuantumMaxwellsdemona, kim2011Quantumszilardengine, cai2012MultiparticlequantumSzilard, parrondo2015Thermodynamicsinformation}. Early theoretical models of these engines typically assumed ideal state discrimination by the demon. More recent studies have moved beyond this
idealization to explore the thermodynamic consequences of measurement errors and non-ideal information processing \citep{wachtler2016Stochasticthermodynamicsbased, still2020ThermodynamicCostBenefit,anderson2022GeneralizedLandauerBound, saha2023InformationEngineNonequilibrium, zhou2024Finitetimeoptimizationquantum}.

For non-orthogonal states, however, discrimination errors are not merely
practical limitations: they are unavoidable consequences of quantum mechanics and are closely tied to thermodynamic consistency. Peres revealed that a demon capable of perfectly distinguishing such states could extract work from a single heat reservoir, directly violating the Kelvin-Planck statement of the second law of thermodynamics
\citep{peres1997Quantumtheoryconcepts,maruyama2009ColloquiumphysicsMaxwells}. More recently, Polo-G\'{o}mez showed that the quantum-mechanical bound on state discrimination \citep{bae2013Structureminimumerrorquantum,bae2015Quantumstatediscrimination} is sufficient to preserve the second law, while the second law itself yields a weaker thermodynamic bound on the achievable discrimination accuracy \citep{polo-gomez2024Thermodynamicboundquantum}. These results establish a direct connection between quantum indistinguishability and thermodynamic
consistency, but the existing bounds are formulated for single-reservoir scenarios. While the second law forbids work extraction in such isothermal environments, Carnot's theorem permits it when a thermal gradient between multiple reservoirs is available~\citep{carnot1872Reflexionspuissancemotrice}. This immediately raises a complementary
question: if a thermal gradient enables work extraction, what fundamental limits govern a thermal machine whose operation relies on the discrimination of non-orthogonal states? Addressing this question requires understanding how microscopic quantum indistinguishability and macroscopic thermal driving jointly constrain energy conversion.

In this Letter, we propose a two-reservoir information machine to map these constraints explicitly. By utilizing a Maxwell's demon to perform imperfect discrimination of the non-orthogonal states during isothermal expansion, the machine directly converts quantum information into mechanical work. The competition between the quantum upper bound on discrimination accuracy and the thermodynamic lower bound required for positive work defines distinct operational regimes in the $(\eta_C,\mu)$ plane. Most strikingly, under weak thermal driving, increasing the state overlap can first suppress positive-work operation and then restore it as the states become nearly identical. This reentrant behavior provides a direct operational manifestation of the competition between quantum indistinguishability and macroscopic thermal driving.
\begin{figure*}
    \centering
 \includegraphics[width=1\linewidth]{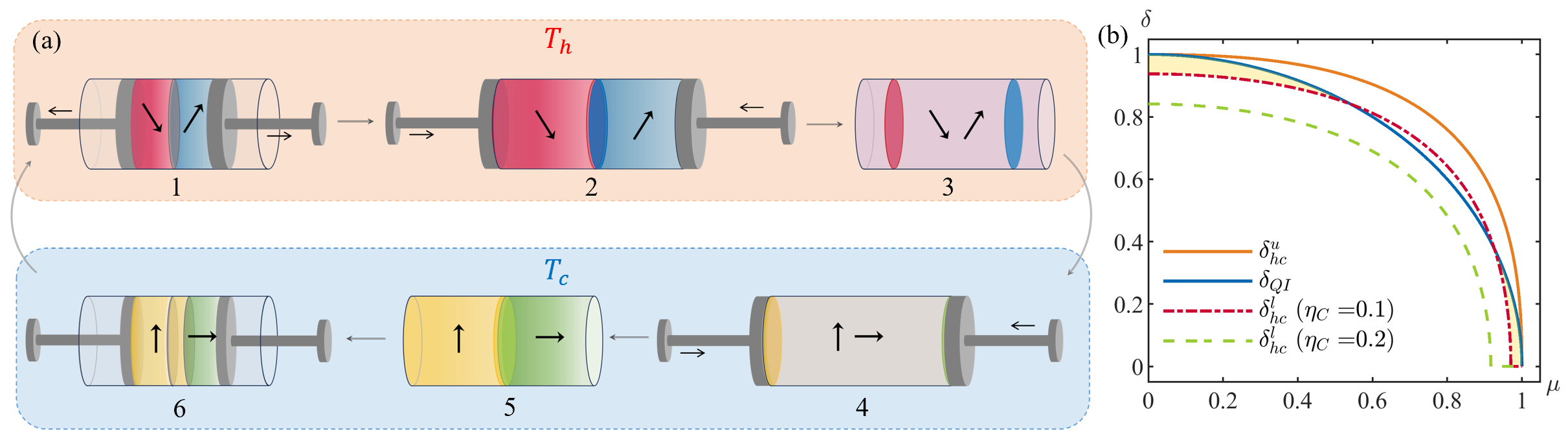} 
   \caption{(a) Schematic of the quantum information thermodynamic cycle. The cycle comprises isothermal expansion (hot reservoir, steps 1-3) and isothermal compression (cold reservoir, steps 4-6), driven by non-orthogonal state discrimination. Red/blue regions denote gas in non-orthogonal states $|\psi_1\rangle$ ($\searrow$) and $|\psi_2\rangle$ ($\nearrow$); yellow/green regions denote orthogonal states $|\phi_1\rangle$ ($\uparrow$) and $|\phi_2\rangle$ ($\rightarrow$). (b) Parameter space for positive work extraction. Discrimination accuracy $\delta$ is plotted against state overlap $\mu$. Functional operation requires $\delta$ to fall between the quantum information limit $\delta_{QI}$ (solid blue curve) and the thermodynamic lower bound $\delta_{hc}^l$ (dashed curves). The shaded yellow area highlights the positive net work regime for $\eta_C = 0.1$.}
    \label{fig:cycle}
\end{figure*}

\textit{Operation of the quantum machine}.---We consider an ideal gas comprising $N$ particles, each possessing an internal quantum degree of freedom completely decoupled from its spatial kinematics. As illustrated in Fig.~\ref{fig:cycle}(a), the gas is initially bisected into two equal portions, prepared in non-orthogonal pure states $|\psi_1\rangle$ and $|\psi_2\rangle$, respectively. Each portion initially occupies a quarter of the total cylinder volume. The degree of indistinguishability between these two states is quantified by their overlap, $\mu= |\langle\psi_1|\psi_2\rangle| = \cos\theta$. The quantum thermal machine operates between a hot reservoir at temperature $T_h$ and a cold reservoir at temperature $T_c$. The cycle extracts work through the discrimination of non-orthogonal states by an external demon in six steps, as shown in Fig.~\ref{fig:cycle}(a). In step 1, while coupled to $T_h$, the ideal gas undergoes quasistatic isothermal expansion from its initial volume, yielding work~\citep{polo-gomez2024Thermodynamicboundquantum}
\begin{equation}
W_{1\to2} = N k_B T_h \ln 2,
\label{eq:W1}
\end{equation}

In step 2, the central wall is replaced by two movable semi-permeable membranes. The red (blue) membrane is transparent to $|\psi_1\rangle$ ($|\psi _2\rangle$) and opaque to $|\psi _2\rangle$  ($|\psi _1\rangle$).
Subsequently, the demon attempts to distinguish $|\psi_1\rangle$ from $|\psi_2\rangle$ with a probability of success $p_s = (1+\delta)/2$, where $\delta \in [0, 1]$ encapsulates the measurement accuracy. Since the discrimination is imperfect, a fraction $(1-\delta)/2$ of the gas particles that were in $|\psi_1\rangle~(|\psi_2\rangle)$ are misidentified as being in  $|\psi_2\rangle~(|\psi_1\rangle)$ and exert pressure on the incorrect membrane~\citep{polo-gomez2024Thermodynamicboundquantum}. Mechanical equilibrium is established when the pressures on both sides of the membrane equalize, which strictly requires the expansion volume fraction $f$ to perfectly match the misidentified particle fraction $f=(1-\delta)/2$. The combined work of measurement and this hindered expansion is~\citep{polo-gomez2024Thermodynamicboundquantum}
\begin{equation}
W_{2\to3} = N k_B T_h \ln 2 \left[1 - H\left(\frac{1+\delta}{2}\right)\right]+W_{\mathrm{meas}}.
\label{eq:W2}
\end{equation}
where $H(p) = -p \log_2 p - (1-p) \log_2(1-p)$ is the the binary Shannon entropy~\citep{shannon1948Mathematicaltheorycommunication}, which quantifies the thermodynamic penalty of the irreversible mixing caused by misidentified particles expanding into the incorrect volume fraction. 

In step 3, the demon's memory is erased and the internal gas states are reset, with an associated work cost $W_{3\to4}=W_{\mathrm{reset}}$. Under the assumption that this joint measurement-resetting process operates reversibly, i.e., $W_{\mathrm{meas}}+W_{\mathrm{reset}}=0$~\citep{polo-gomez2024Thermodynamicboundquantum}. The total output work extracted from the hot reservoir simplifies to
\begin{equation}
W_{out} =\sum_{i=1}^{i=3} W_{i\to i+1}= N k_B T_h \ln 2 \left[2 - H\left(\frac{1+\delta}{2}\right)\right],
\label{eq:Wout}
\end{equation}
which is entirely supplied by the heat absorption from the hot reservoir, namely, $Q_h=W_{out}>0$. Here, the acquired information about the quantum states is actively converted into mechanical work, embodying the quintessential mechanism of an information engine. Tracing out the spatial degrees of freedom, the gas is described by a mixed state, which can be diagonalized into an orthogonal basis
\begin{equation}
\hat{\rho} = \frac{1}{2}\sum_{i=1,2}|\psi_i\rangle\langle\psi_i| = c|\phi_1\rangle\langle\phi_1| + (1-c)|\phi_2\rangle\langle\phi_2|,
\label{eq:rho}
\end{equation}
where the real constant $c\in[0,1]$. The cycle then couples to the cold reservoir. In step 4, two semipermeable membranes, the yellow and green membranes in Fig.~\ref{fig:cycle}(a), are introduced to mechanically separate the orthogonal components. Because orthogonal states can be perfectly discriminated, separating these components requires work
\begin{equation}
W_{4\to5} = N k_B T_c \ln 2.
\label{eq:W3}
\end{equation}

In step 5, isothermal compression restores the system to its initial thermodynamic pressure. Since the separated orthogonal components contain particle fractions $c$ and $1-c$, isothermally compressing them to their respective initial partial densities requires work
\begin{equation}
W_{5\to6} = N k_B T_c \ln 2 H(c).
\label{eq:W4}
\end{equation}
Finally, in step 6, partitions are inserted and quasistatic unitary transformations reversibly restore the initial non-orthogonal states at zero work cost ($W_{6\to 1} = 0$), closing the cycle. Summing Eqs.~(\ref{eq:W3}) and (\ref{eq:W4}), the total work performed on the system during compression is
\begin{equation}
W_{in} = W_{4\to 5}+W_ {5\to6}= N k_B T_c \ln 2 \left[1 + H\left(c\right)\right],
\label{eq:Win}
\end{equation}
where $c = (1+\mu)/2$~\citep{polo-gomez2024Thermodynamicboundquantum}, and $H(c)$ represents the von Neumann entropy of the mixed state. Crucially, this $H(c)$ term embodies the Landauer erasure cost required to reset the correlations of the non-orthogonal mixture. The total heat rejected to the cold reservoir is given by $Q_c=-W_{in}<0$. This step explicitly maps the thermodynamic penalty associated with
the nonorthogonality of the initial states into a macroscopic work cost.

\textit{Work output and efficiency of the machine}.--- The net work extracted over one full cycle is defined as
\begin{equation}
W_{net} = W_{out} - W_{in}.
\label{eq:Wnet}
\end{equation}
For a two-reservoir information machine with a thermal bias $\eta_C = 1 - T_c/T_h > 0$, We characterize its performance by efficiency $\eta \equiv W_{net}/Q_h=W_{net}/W_{out}$. It follows from Eqs.~\eqref{eq:Wout} and \eqref{eq:Win} that
\begin{equation}
\eta = 1 - (1-\eta_C)\frac{1+H(\frac{1+\mu}{2})}{2-H\left(\frac{1+\delta}{2}\right)}.
\label{eq:efficiency}
\end{equation}
According to the Second Law of Thermodynamics, the efficiency is strictly bounded by Carnot's theorem, dictating $\eta \le \eta_C$ which directly yields
\begin{equation}
H\left(\frac{1+\delta}{2}\right)+H\left(\frac{1+\mu}{2}\right) \ge 1 .
\label{eq:thermo_upper}
\end{equation}
Since the Shannon entropy $H(x)$ decreases monotonically for $x \in [0.5, 1]$, this inequality imposes a strict thermodynamic upper bound on the discrimination accuracy, denoted as $\delta_{hc}^u$. Notably, this bound recovers the result obtained in single-reservoir scenario~\citep{polo-gomez2024Thermodynamicboundquantum}, demonstrating its universality. 

In addition to the thermodynamic constraint, the foundational principles of quantum mechanics inherently restrict the maximum achievable discrimination accuracy. Governed by the Holevo-Helstrom theorem~\citep{helstrom1969Quantumdetectionestimation, holevo1974remarks, helstrom1976quantum}, the optimal probability of correctly identifying the states translates into a quantum information upper bound for the discrimination accuracy,
\begin{equation}
\delta \le \delta_{QI} = \sin\theta = \sqrt{1-\mu^2}.
\label{eq:QI_bound}
\end{equation}
A physically permissible machine must simultaneously satisfy the restrictions imposed by both thermodynamics and quantum mechanics. However, as previously demonstrated~\citep{polo-gomez2024Thermodynamicboundquantum}, the quantum bound [blue solid curve in Fig. \ref{fig:cycle}(b)] provides a strictly tighter restriction than the thermodynamic upper bound [orange solid curve in Fig. \ref{fig:cycle}(b)], namely, $\delta_{QI} < \delta_{hc}^u$. This indicates that quantum indistinguishability, rather than the thermodynamic bound derived from the second law, represents the ultimate theoretical ceiling for state discrimination.

While quantum mechanics sets the fundamental ceiling for discrimination accuracy, thermodynamics imposes a critical floor for practical energy extraction. To operate as a functional heat engine, the system must output positive net work ($\eta > 0$). Applying this operational condition to Eq.~(\ref{eq:efficiency}) yields
\begin{equation}
H\left(\frac{1+\delta}{2}\right) < 2 - (1-\eta_C)\left[1+H\left(\frac{1+\mu}{2}\right)\right],
\label{eq:thermo_lower}
\end{equation}
which defines a critical thermodynamic lower bound, $\delta > \delta_{hc}^l(\eta_C, \mu)$. Therefore, the parameter space capable of positive work extraction is strictly confined to the intersection region $\delta_{hc}^l < \delta \le \delta_{QI}$. As illustrated in Fig.~\ref{fig:cycle}(b), this interplay reveals a striking physical consequence: the functional region splits into two disconnected segments along the non-orthogonality axis, $0 < \mu < \mu_1$ and $\mu_2 < \mu < 1$. Within the intermediate gap $\mu_1 \le \mu \le \mu_2$, although state discrimination operates well within the limits permitted by quantum mechanics, the accuracy remains insufficient to overcome the thermodynamic lower bound $\delta_{hc}^l$, fundamentally prohibiting the machine from outputting positive work.

\begin{figure}
    \centering
    \includegraphics[width=\columnwidth]{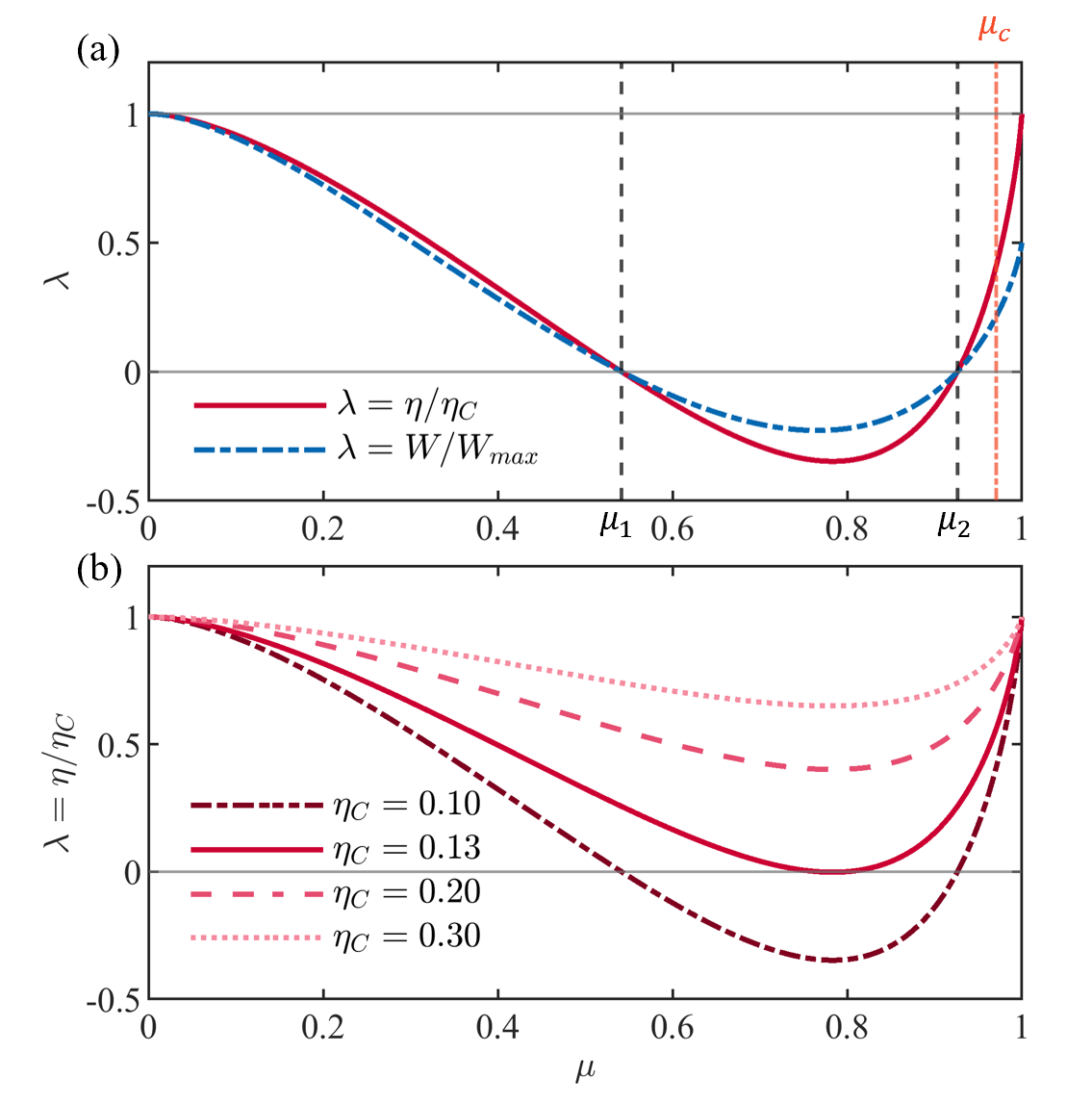} 
    \caption{Performance characteristics of the machine evaluated at $\delta = \delta_{QI}$. (a) The normalized net work output and efficiency are plotted as functions of $\mu$ for $\eta_C = 0.1$. (b) Normalized efficiency as a function of $\mu$ with different $\eta_C $. }
    \label{fig:work_efficiency}
\end{figure}

As illustrated in Fig.~\ref{fig:work_efficiency}, tracing the cycle at the optimal quantum limit ($\delta = \delta_{QI}$) reveals that the net work and efficiency exhibit strongly non-monotonic behavior with respect to the state overlap $\mu$. Notably, the normalized efficiency uniformly exhibits dual asymptotes, recovering the Carnot limit ($\eta/\eta_C \to 1$) at both ideal extremes ($\mu \to 0$ and $\mu \to 1$), despite different physical mechanisms. 
i) For orthogonal states ($\mu\to0$), the demon achieves perfect distinguishability ($\delta_{QI} \to 1$), eliminating any error-induced pressure imbalance during expansion ($H[(1+\delta_{QI})/2]\to0$). The machine effectively operates as two distinct, perfectly resolved gas components, extracting the absolute maximum expansion work. Simultaneously, the completely orthogonal mixture represents a maximum classical entropy ($H(c)\to1$), requiring substantial but justifiable reset work. This optimal information-to-work conversion yields the global maximum net work $W_{net}=2Nk_BT_h\eta_C\ln{2}$. 
ii) Conversely, when the states are identical ($\mu\to1$), they become fundamentally indistinguishable ($\delta_{QI} \to 0$). The demon is forced to make random guesses, resulting in a complete loss of directional information that nullifies the second-stage expansion work ($H[(1+\delta_{QI})/2]\to1$). However, precisely because the states are identical, the post-measurement ensemble is effectively a pure state with zero von Neumann entropy ($H(c)\to0$). This vanishing entropy dictates that the memory erasure and reset stage requires minimal compression work, entirely offsetting the effect of the reduced expansion output on the efficiency. Physically, the indistinguishable mixture degenerates into a single, unified ideal gas undergoing a standard Carnot-like cycle. Consequently, the net work evaluates to $W_{net}=Nk_BT_h\eta_C\ln{2}$, exactly one half of the maximum work, simply because the machine is now effectively cycling one unified gas portion instead of exploiting two perfectly resolvable ones.

Moving away from these ideal extremes, the machine's performance is penalized by the combined informational cost. In the intermediate-overlap region, the states remain sufficiently distinct to retain a substantial Landauer erasure cost $H(c)$, while their nonorthogonality simultaneously limits the extractable expansion work through discrimination errors. Consequently, the reset cost can exceed the expansion work, driving the net work into the negative regime between $\mu_1$ and $\mu_2$ and rendering the machine dissipative. The severity of this informational penalty, however, is highly sensitive to the available macroscopic thermal resources. As explicitly shown in Fig.~\ref{fig:work_efficiency}(b), increasing the thermal gradient $\eta_C$ systematically lifts the efficiency curve. A stronger thermal drive provides sufficient expansion work to offset the erasure cost, progressively narrowing the negative-work gap until it eventually disappears entirely. 

When the applied thermal gradient is large enough, the thermodynamic lower restriction vanishes ($\delta_{hc}^l \to 0$). By setting $\delta = 0$ in the left side of Eq.~\eqref{eq:thermo_lower}, we analytically identify a critical cutoff 
\begin{equation}
\mu_c = 2H^{-1}\left(\frac{\eta_C}{1-\eta_C}\right) - 1.
\label{eq:cutoff}
\end{equation}
For highly non-orthogonal states ($\mu > \mu_c$), the thermodynamic penalty of resetting the coherent mixed state is entirely outweighed by the thermal gradient, enabling unconditional positive work output for any quantum-mechanically allowed $\delta$. In the limit of a vanishing thermal bias ($\eta_C \to 0$), the entropy argument requires $H\left[(1+\mu_c)/2\right] \approx \eta_C$. Utilizing the asymptotic expansion of the Shannon entropy near certainty, the cutoff boundary approaches unity logarithmically as $\mu_c \sim 1 - \mathcal{O}\left[\eta_C/\ln(1/\eta_C)\right]$. This implies that for extremely small temperature gradients, only nearly identical states can overcome the prohibitive memory erasure penalty. Conversely, when $\eta_C \ge 0.5$, the ratio $\eta_C/(1-\eta_C) \ge 1$, which yields $\mu_c = 0$. Remarkably, for any robust thermal gradient $\eta_C>0.5$, the macroscopic classical resource completely overpowers the microscopic quantum erasure cost, rendering the thermodynamic restriction obsolete across all possible state overlaps.

\begin{figure}[htb!]
    \centering
    \includegraphics[width=\linewidth]{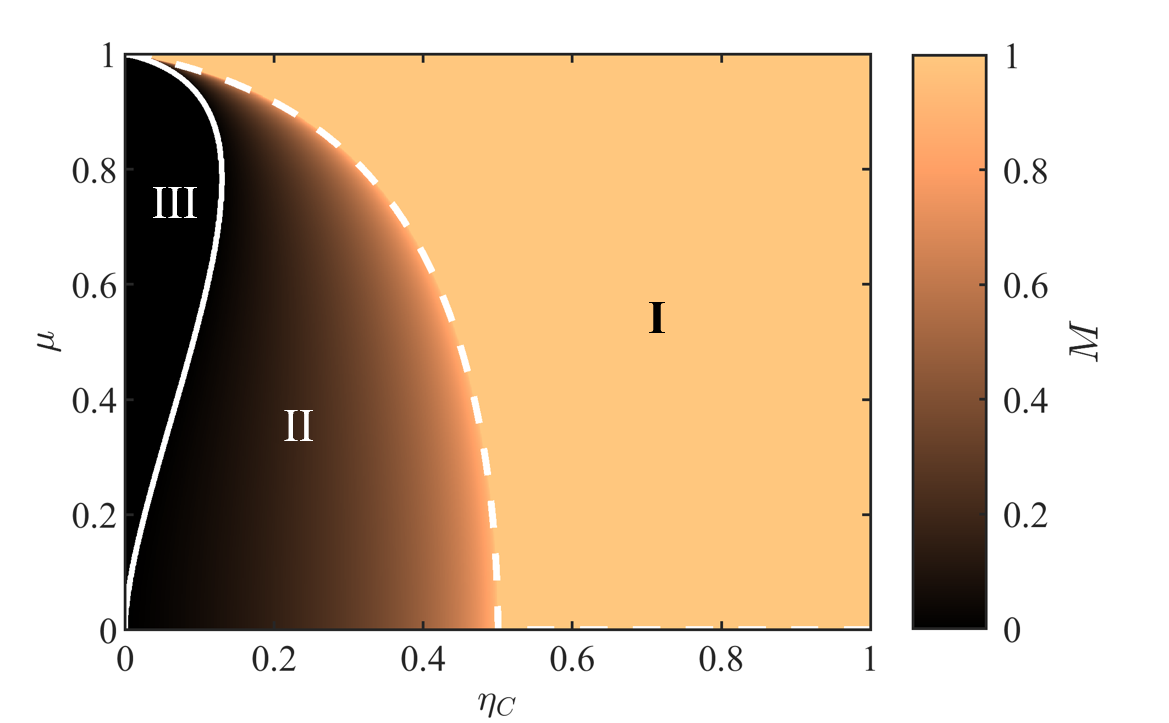} 
    \caption{Functional phase diagram of the quantum information machine in the parameter space of Carnot efficiency $\eta_C$ versus state overlap $\mu$. The color map indicates the normalized order parameter $M$. The diagram features three distinct regimes: I. the Pure Heat Engine Phase ($M=1$), II. the Mixed Phase ($0<M<1$), and III. the Dissipative Phase ($M=0$). The dashed white line denotes the analytical boundary $\mu_c$ given by Eq.~(\ref{eq:cutoff}).}
    \label{fig:phasediagram}
\end{figure}

\textit{Phase diagram of the machine}.---To systematically map the thermodynamic consequences of quantum indistinguishability, we classify the machine's emergent functioning modes with a normalized order parameter
\begin{equation}
M = \frac{\delta_{QI} - \delta_{hc}^l}{\delta_{QI}} \cdot \Theta(W_{net}),
\label{eq:order_parameter}
\end{equation}
where $\Theta(\cdot)$ is the Heaviside step function. Because physical measurements require $0 \le \delta \le \delta_{QI}$, $M$ strictly measures the fraction of the realizable quantum parameter space that permits positive work extraction ($\delta_{hc}^l < \delta \le \delta_{QI}$). As depicted in the phase diagram [Fig.~\ref{fig:phasediagram}], continuously tuning the state overlap $\mu$ and the thermal gradient (associated with $\eta_C$) drives the system through three distinct operational phases: the Pure Heat Engine phase (I), the Mixed phase (II), and the Dissipative phase (III).

The distinct phases reflect the competition between the energetic cost of information erasure and the available thermodynamic driving force. In the \textit{Pure Heat Engine Phase I} ($M = 1$), the robust thermal gradient completely eclipses the erasure cost ($\mu > \mu_c$, $\delta_{hc}^l = 0$), guaranteeing unconditional $W_{net} > 0$ for any physically allowed $\delta$. The \textit{Mixed Phase II} ($0 < M < 1$) represents a fragile regime where operation becomes highly protocol-sensitive. Here, the machine functions as either a valid engine or a dissipative heater depending on the specific realization of $\delta$. Finally, the \textit{Dissipative Phase III} ($M=0$) maps exactly to the intermediate failure gap ($\mu_1 \le \mu \le \mu_2$) identified in Fig.~\ref{fig:work_efficiency}. Here, the required entropy reduction severely overpowers expansion work, strictly prohibiting positive work generation.

Globally, the phase diagram partitions into three defining regimes governed by the thermal gradient. For large gradients ($\eta_C \ge 0.5$), the macroscopic thermal resource absolutely prevails ($\mu_c = 0$). The Dissipative phase vanishes entirely, ensuring positive work output across all state overlaps. For intermediate gradients ($0.13 < \eta_C < 0.5$), a dominant physical constraint crossover occurs across the analytical boundary $\mu_c$ (white dashed line). Most strikingly, in the weak-gradient regime ($\eta_C \lesssim 0.13$), the system exhibits am anomalous \textit{reentrant phase transition}. As $\mu$ increases, the machine originates in the Mixed phase, plunges into the Dissipative phase, re-emerges into the Mixed phase, and ultimately stabilizes in the Pure Heat Engine phase. This reentrant behavior encapsulates the linear scaling of $\delta_{QI}$ clashing against the highly nonlinear $\delta_{hc}^l$: moderate overlap induces a catastrophic entropy penalty, whereas near-perfect overlap restores functionality due to the vanishing mixture entropy $H(c) \to 0$.

The disparate visual boundaries in the phase diagram are dictated by the mathematical singularities of the underlying physical bounds. The transition across $\mu_c$ is remarkably abrupt. Because $\partial H(p)/\partial p |_{p=1/2} = 0$, the entropy behaves quadratically near $\delta_{hc}^l \to 0$. Consequently, the thermodynamic bound exhibits a square-root singularity, $|\partial \delta_{hc}^l / \partial \mu| \to \infty$. This divergent sensitivity dictates that a minuscule decrease in state distinguishability across the $\mu_c$ line abruptly triggers a massive thermodynamic penalty, visually manifesting as a steep, pseudo-first-order color gradient. Conversely, the continuous boundary $\delta_{hc}^l = \delta_{QI}$ (solid white line) separating regions II and III features finite entropy derivatives because $\delta_{QI}$ strictly resides within $(0,1)$. The valid parameter space depletes smoothly and linearly with respect to $\mu$, generating the gradual fading ($M \to 0$) into the fully forbidden dark region.

\textit{Concluding remarks}.---In this Letter, we investigated how microscopic quantum indistinguishability and macroscopic thermodynamic constraints jointly govern information-to-energy conversion. By mapping the functional phase diagram of a two-reservoir information machine, we demonstrated that its operational boundaries emerge directly from the competition between the thermal gradient and the Landauer erasure cost of non-orthogonal states. Notably, under weak thermal driving, this interplay induces an anomalous reentrant transition of operational regimes: As state overlap increases, the machine plunges into a purely dissipative regime before near-perfect indistinguishability unexpectedly restores its capacity to output positive work.

Fundamentally, our results endow the abstract Holevo-Helstrom bound with a concrete operational interpretation, showing how the
quantum constraint on state discrimination safeguards consistency with the Second Law in the present information-machine setting. Non-orthogonal state discrimination inherently generates entropy, and apparent Second Law violations stem from neglecting the thermodynamic cost of its inverse process, nonorthogonal-state erasure (NOSE)~\citep{xia}. The competition mapped in our phase diagram is precisely the thermal gradient overcoming this intrinsic NOSE penalty. The gap ($\delta_{QI} < \delta_{hc}^u$) physically reflects this unitarity-mandated entropy increase, naturally safeguarding the Second Law of Thermodynamics.

Building on this framework, future studies could introduce multi-partite entangled working substances to explore how non-local correlations might alter these operational boundaries and mitigate classical limits~\citep{cai2012MultiparticlequantumSzilard,kamimura2022QuantumEnhancedHeatEngine,bai2025Multipartiteentanglementmeasures}. Furthermore, advancing this static framework into finite-time thermodynamics~\citep{kosloff2014quantum,Bjarne2022,qiu20roadmap} will illuminate how the underlying dynamics of quantum information processing and non-equilibrium thermal processes~\citep{2022Ma,2013PRLMQJ,Fadler2023,zhou2024Finitetimeoptimizationquantum,chen2026finite}  mutually restrict fast energy extraction. Ultimately, connecting quantum distinguishability with Carnot efficiency lays a rigorous foundation for designing the next generation of high-performance quantum thermal devices.

\textit{Acknowledgment---.}This work is supported by the National Natural Science Foundation of China under Grant No. 12688201 and Science Challenge Project under Grant No.TZ2025017. Y.-H.Ma thanks the National Natural Science Foundation of China for support under Grants No. 12305037 and No. 12675036
\end{CJK*}

\bibliography{Refs}

@article{alhambra2016FluctuatingWorkQuantum,
  title = {Fluctuating {{Work}}: {{From Quantum Thermodynamical Identities}} to a {{Second Law Equality}}},
  shorttitle = {Fluctuating {{Work}}},
  author = {Alhambra, {\'A}lvaro M. and Masanes, Lluis and Oppenheim, Jonathan and Perry, Christopher},
  year = 2016,
  month = oct,
  journal = {Phys. Rev. X},
  volume = {6},
  number = {4},
  pages = {041017},
  doi = {10.1103/PhysRevX.6.041017}
}

@article{alicki2004ThermodynamicsQuantumInformation,
  title = {Thermodynamics of {{Quantum Information Systems}} --- {{Hamiltonian Description}}},
  author = {Alicki, Robert and Horodecki, Micha{\l} and Horodecki, Pawe{\l} and Horodecki, Ryszard},
  year = 2004,
  month = sep,
  journal = {Open Syst. Inf. Dyn.},
  volume = {11},
  number = {03},
  pages = {205--217},
  doi = {10.1023/B:OPSY.0000047566.72717.71}
}

@article{anderson2022GeneralizedLandauerBound,
  title = {Generalized {{Landauer Bound}} for {{Information Processing}}: {{Proof}} and {{Applications}}},
  shorttitle = {Generalized {{Landauer Bound}} for {{Information Processing}}},
  author = {Anderson, Neal G.},
  year = 2022,
  month = oct,
  journal = {Entropy},
  volume = {24},
  number = {11},
  pages = {1568},
  doi = {10.3390/e24111568}
}

@article{bae2013Structureminimumerrorquantum,
  title = {Structure of Minimum-Error Quantum State Discrimination},
  author = {Bae, Joonwoo},
  year = 2013,
  month = jul,
  journal = {New J. Phys.},
  volume = {15},
  number = {7},
  pages = {073037},
  doi = {10.1088/1367-2630/15/7/073037}
}

@article{bae2015Quantumstatediscrimination,
  title = {Quantum State Discrimination and Its Applications},
  author = {Bae, Joonwoo and Kwek, Leong-Chuan},
  year = 2015,
  month = feb,
  journal = {J. Phys. A: Math. Theor.},
  volume = {48},
  number = {8},
  pages = {083001},
  doi = {10.1088/1751-8113/48/8/083001}
}

@article{bai2025Multipartiteentanglementmeasures,
  title = {Multipartite Entanglement Measures Based on the Thermodynamic Framework},
  author = {Bai, Chen-Ming and Luo, Yu},
  year = 2025,
  month = sep,
  journal = {Phys. Rev. A},
  volume = {112},
  number = {3},
  pages = {032424},
  doi = {10.1103/rg65-sh1f}
}

@article{bennett1982Thermodynamicscomputationreview,
  title = {The Thermodynamics of Computation---a Review},
  author = {Bennett, Charles H.},
  year = 1982,
  month = dec,
  journal = {Int. J. Theor. Phys.},
  volume = {21},
  number = {12},
  pages = {905--940},
  doi = {10.1007/BF02084158}
}

@article{cai2012MultiparticlequantumSzilard,
  title = {Multiparticle Quantum {{Szilard}} Engine with Optimal Cycles Assisted by a {{Maxwell}}'s Demon},
  author = {Cai, C. Y. and Dong, H. and Sun, C. P.},
  year = 2012,
  month = mar,
  journal = {Phys. Rev. E},
  volume = {85},
  number = {3},
  pages = {031114},
  doi = {10.1103/PhysRevE.85.031114}
}

@article{carnot1872Reflexionspuissancemotrice,
  title = {R\'eflexions Sur La Puissance Motrice Du Feu et Sur Les Machines Propres \`a D\'evelopper Cette Puissance},
  author = {Carnot, S.},
  year = 1872,
  journal = {Ann. Sci. \'Ecole Norm. Sup.},
  volume = {1},
  pages = {393--457},
  doi = {10.24033/asens.88}
}

@article{dong2011QuantumMaxwellsdemona,
  title = {Quantum {{Maxwell}}'s Demon in Thermodynamic Cycles},
  author = {Dong, H. and Xu, D. Z. and Cai, C. Y. and Sun, C. P.},
  year = 2011,
  month = jun,
  journal = {Phys. Rev. E},
  volume = {83},
  number = {6},
  pages = {061108},
  doi = {10.1103/PhysRevE.83.061108}
}

@article{esposito2011SecondlawLandauer,
  title = {Second Law and {{Landauer}} Principle Far from Equilibrium},
  author = {Esposito, M. and Van Den Broeck, C.},
  year = 2011,
  month = aug,
  journal = {EPL},
  volume = {95},
  number = {4},
  pages = {40004},
  doi = {10.1209/0295-5075/95/40004}
}

@article{helstrom1969Quantumdetectionestimation,
  title = {Quantum Detection and Estimation Theory},
  author = {Helstrom, Carl W.},
  year = 1969,
  journal = {J Stat Phys},
  volume = {1},
  number = {2},
  pages = {231--252},
  doi = {10.1007/BF01007479}
}

@book{helstrom1976quantum,
  title = {Quantum Detection and Estimation Theory},
  author = {Helstrom, Carl W.},
  year = 1976,
  publisher = {Academic Press},
  address = {New York}
}

@article{hilt2011Landauersprinciplequantum,
  title = {Landauer's Principle in the Quantum Regime},
  author = {Hilt, Stefanie and Shabbir, Saroosh and Anders, Janet and Lutz, Eric},
  year = 2011,
  month = mar,
  journal = {Phys. Rev. E},
  volume = {83},
  number = {3},
  pages = {030102},
  doi = {10.1103/PhysRevE.83.030102}
}

@article{holevo1974remarks,
  title = {Remarks on Optimal Quantum Measurements},
  author = {Holevo, Alexander S},
  year = 1974,
  journal = {Probl. Inf. Transm.},
  volume = {10},
  number = {4},
  pages = {317--320}
}

@article{jacobs2009Secondlawthermodynamics,
  title = {Second Law of Thermodynamics and Quantum Feedback Control: {{Maxwell}}'s Demon with Weak Measurements},
  shorttitle = {Second Law of Thermodynamics and Quantum Feedback Control},
  author = {Jacobs, Kurt},
  year = 2009,
  month = jul,
  journal = {Phys. Rev. A},
  volume = {80},
  number = {1},
  pages = {012322},
  doi = {10.1103/PhysRevA.80.012322}
}

@article{kamimura2022QuantumEnhancedHeatEngine,
  title = {Quantum-{{Enhanced Heat Engine Based}} on {{Superabsorption}}},
  author = {Kamimura, Shunsuke and Hakoshima, Hideaki and Matsuzaki, Yuichiro and Yoshida, Kyo and Tokura, Yasuhiro},
  year = 2022,
  month = may,
  journal = {Phys. Rev. Lett.},
  volume = {128},
  number = {18},
  pages = {180602},
  doi = {10.1103/PhysRevLett.128.180602}
}

@article{kim2011QuantumSzilardEngine,
  title = {Quantum {{Szilard Engine}}},
  author = {Kim, Sang Wook and Sagawa, Takahiro and De Liberato, Simone and Ueda, Masahito},
  year = 2011,
  month = feb,
  journal = {Phys. Rev. Lett.},
  volume = {106},
  number = {7},
  pages = {070401},
  doi = {10.1103/PhysRevLett.106.070401}
}

@article{landauer1961Irreversibilityheatgeneration,
  title = {Irreversibility and Heat Generation in the Computing Process},
  author = {Landauer, R.},
  year = 1961,
  journal = {IBM J. Res. Dev.},
  volume = {5},
  number = {3},
  pages = {183--191},
  doi = {10.1147/rd.53.0183}
}

@article{maruyama2009ColloquiumphysicsMaxwells,
  title = {{\emph{Colloquium}} : {{The}} Physics of {{Maxwell}}'s Demon and Information},
  shorttitle = {{\emph{Colloquium}}},
  author = {Maruyama, Koji and Nori, Franco and Vedral, Vlatko},
  year = 2009,
  month = jan,
  journal = {Rev. Mod. Phys.},
  volume = {81},
  number = {1},
  pages = {1--23},
  doi = {10.1103/RevModPhys.81.1}
}

@article{naghiloo2018InformationGainLoss,
  title = {Information {{Gain}} and {{Loss}} for a {{Quantum Maxwell}}'s {{Demon}}},
  author = {Naghiloo, M. and Alonso, J. J. and Romito, A. and Lutz, E. and Murch, K. W.},
  year = 2018,
  month = jul,
  journal = {Phys. Rev. Lett.},
  volume = {121},
  number = {3},
  pages = {030604},
  doi = {10.1103/PhysRevLett.121.030604}
}

@article{parrondo2015Thermodynamicsinformation,
  title = {Thermodynamics of Information},
  author = {Parrondo, Juan MR and Horowitz, Jordan M and Sagawa, Takahiro},
  year = 2015,
  journal = {Nat. Phys.},
  volume = {11},
  number = {2},
  pages = {131--139},
  publisher = {Nature Publishing Group},
  doi = {10.1038/nphys3230}
}

@book{peres1997Quantumtheoryconcepts,
  title = {Quantum Theory: Concepts and Methods},
  author = {Peres, Asher},
  year = 1997,
  volume = {57},
  pages = {260--297},
  publisher = {Springer}
}

@article{polo-gomez2024Thermodynamicboundquantum,
  title = {Thermodynamic Bound on Quantum State Discrimination},
  author = {{Polo-G{\'o}mez}, Jos{\'e}},
  year = 2024,
  month = jan,
  journal = {Phys. Rev. E},
  volume = {109},
  number = {1},
  pages = {014119},
  doi = {10.1103/PhysRevE.109.014119}
}

@article{ptaszynski2019ThermodynamicsQuantumInformation,
  title = {Thermodynamics of {{Quantum Information Flows}}},
  author = {Ptaszy{\'n}ski, Krzysztof and Esposito, Massimiliano},
  year = 2019,
  month = apr,
  journal = {Phys. Rev. Lett.},
  volume = {122},
  number = {15},
  pages = {150603},
  doi = {10.1103/PhysRevLett.122.150603}
}

@article{quan2006MaxwellsDemonAssisted,
  title = {Maxwell's {{Demon Assisted Thermodynamic Cycle}} in {{Superconducting Quantum Circuits}}},
  author = {Quan, H. T. and Wang, Y. D. and Liu, Yu-xi and Sun, C. P. and Nori, Franco},
  year = 2006,
  month = oct,
  journal = {Phys. Rev. Lett.},
  volume = {97},
  number = {18},
  pages = {180402},
  doi = {10.1103/PhysRevLett.97.180402}
}

@article{reeb2014ImprovedLandauerprinciple,
  title = {An Improved {{Landauer}} Principle with Finite-Size Corrections},
  author = {Reeb, David and Wolf, Michael M},
  year = 2014,
  month = oct,
  journal = {New J. Phys.},
  volume = {16},
  number = {10},
  pages = {103011},
  publisher = {IOP Publishing},
  doi = {10.1088/1367-2630/16/10/103011}
}

@article{sagawa2009MinimalEnergyCost,
  title = {Minimal {{Energy Cost}} for {{Thermodynamic Information Processing}}: {{Measurement}} and {{Information Erasure}}},
  shorttitle = {Minimal {{Energy Cost}} for {{Thermodynamic Information Processing}}},
  author = {Sagawa, Takahiro and Ueda, Masahito},
  year = 2009,
  month = jun,
  journal = {Phys. Rev. Lett.},
  volume = {102},
  number = {25},
  pages = {250602},
  doi = {10.1103/PhysRevLett.102.250602}
}

@article{sagawa2010GeneralizedJarzynskiEquality,
  title = {Generalized {{Jarzynski Equality}} under {{Nonequilibrium Feedback Control}}},
  author = {Sagawa, Takahiro and Ueda, Masahito},
  year = 2010,
  month = mar,
  journal = {Phys. Rev. Lett.},
  volume = {104},
  number = {9},
  pages = {090602},
  doi = {10.1103/PhysRevLett.104.090602}
}

@article{sagawa2012Fluctuationtheoreminformation,
  title = {Fluctuation Theorem with Information Exchange: {{Role}} of Correlations in Stochastic Thermodynamics},
  author = {Sagawa, Takahiro and Ueda, Masahito},
  year = 2012,
  month = nov,
  journal = {Phys. Rev. Lett.},
  volume = {109},
  number = {18},
  pages = {180602},
  publisher = {American Physical Society},
  doi = {10.1103/PhysRevLett.109.180602}
}

@article{saha2023InformationEngineNonequilibrium,
  title = {Information {{Engine}} in a {{Nonequilibrium Bath}}},
  author = {Saha, Tushar K. and Ehrich, Jannik and Gavrilov, Mom{\v c}ilo and Still, Susanne and Sivak, David A. and Bechhoefer, John},
  year = 2023,
  month = aug,
  journal = {Phys. Rev. Lett.},
  volume = {131},
  number = {5},
  pages = {057101},
  doi = {10.1103/PhysRevLett.131.057101}
}

@article{shannon1948Mathematicaltheorycommunication,
  title = {A Mathematical Theory of Communication},
  author = {Shannon, Claude E},
  year = {Jul and Oct 1948},
  journal = {The Bell System Technical Journal},
  volume = {27},
  number = {3},
  pages = {379--423 and 623--656},
  publisher = {Nokia Bell Labs}
}

@article{still2020ThermodynamicCostBenefit,
  title = {Thermodynamic {{Cost}} and {{Benefit}} of {{Memory}}},
  author = {Still, Susanne},
  year = 2020,
  month = feb,
  journal = {Phys. Rev. Lett.},
  volume = {124},
  number = {5},
  pages = {050601},
  doi = {10.1103/PhysRevLett.124.050601}
}

@article{szilard1929UberEntropieverminderungthermodynamischen,
  title = {\"Uber Die {{Entropieverminderung}} in Einem Thermodynamischen {{System}} Bei {{Eingriffen}} Intelligenter {{Wesen}}},
  author = {Szilard, L.},
  year = 1929,
  journal = {Z. Physik},
  volume = {53},
  pages = {840}
}

@article{wachtler2016Stochasticthermodynamicsbased,
  title = {Stochastic Thermodynamics Based on Incomplete Information: Generalized {{Jarzynski}} Equality with Measurement Errors with or without Feedback},
  shorttitle = {Stochastic Thermodynamics Based on Incomplete Information},
  author = {W{\"a}chtler, Christopher W and Strasberg, Philipp and Brandes, Tobias},
  year = 2016,
  month = nov,
  journal = {New J. Phys.},
  volume = {18},
  number = {11},
  pages = {113042},
  doi = {10.1088/1367-2630/18/11/113042}
}

@article{yuen1975Optimumtestingmultiple,
  title = {Optimum Testing of Multiple Hypotheses in Quantum Detection Theory},
  author = {Yuen, H. and Kennedy, R. and Lax, M.},
  year = 1975,
  month = mar,
  journal = {IEEE Trans. Inf. Theory},
  volume = {21},
  number = {2},
  pages = {125--134},
  doi = {10.1109/TIT.1975.1055351}
}

@article{zhou2024Finitetimeoptimizationquantum,
  title = {Finite-Time Optimization of a Quantum {{Szilard}} Heat Engine},
  author = {Zhou, Tan-Ji and Ma, Yu-Han and Sun, C. P.},
  year = 2024,
  month = oct,
  journal = {Phys. Rev. Research},
  volume = {6},
  number = {4},
  pages = {043001},
  doi = {10.1103/PhysRevResearch.6.043001}
}

@Article{Bjarne2022,
AUTHOR = {Andresen, Bjarne and Salamon, Peter},
TITLE = {Future Perspectives of Finite-Time Thermodynamics},
JOURNAL = {Entropy},
VOLUME = {24},
YEAR = {2022},
NUMBER = {5},
ARTICLE-NUMBER = {690},
PubMedID = {35626573},
ISSN = {1099-4300},
DOI = {10.3390/e24050690}
}

@article{kosloff2014quantum,
  title={Quantum heat engines and refrigerators: Continuous devices},
  author={Kosloff, Ronnie and Levy, Amikam},
  journal={Annual review of physical chemistry},
  volume={65},
  number={1},
  pages={365--393},
  year={2014},
  publisher={Annual Reviews},
doi={10.1146/annurev-physchem-040513-103724}
}

@article{2022Ma,
  title = {Minimal energy cost to initialize a bit with tolerable error},
  author = {Ma, Yu-Han and Chen, Jin-Fu and Sun, C. P. and Dong, Hui},
  journal = {Phys. Rev. E},
  volume = {106},
  issue = {3},
  pages = {034112},
  numpages = {8},
  year = {2022},
  month = {Sep},
  publisher = {American Physical Society},
  doi = {10.1103/PhysRevE.106.034112},
  url = {https://link.aps.org/doi/10.1103/PhysRevE.106.034112}
}

@article{2013PRLMQJ,
  title = {Maxwell's Refrigerator: An Exactly Solvable Model},
  author = {Mandal, Dibyendu and Quan, H. T. and Jarzynski, Christopher},
  journal = {Phys. Rev. Lett.},
  volume = {111},
  issue = {3},
  pages = {030602},
  numpages = {5},
  year = {2013},
  month = {Jul},
  publisher = {American Physical Society},
  doi = {10.1103/PhysRevLett.111.030602},
  url = {https://link.aps.org/doi/10.1103/PhysRevLett.111.030602}
}

@article{Fadler2023,
  title = {Efficiency at Maximum Power of a Carnot Quantum Information Engine},
  author = {Fadler, Paul and Friedenberger, Alexander and Lutz, Eric},
  journal = {Phys. Rev. Lett.},
  volume = {130},
  issue = {24},
  pages = {240401},
  numpages = {7},
  year = {2023},
  month = {Jun},
  publisher = {American Physical Society},
  doi = {10.1103/PhysRevLett.130.240401}
}

@article{chen2026finite,
  title={Finite-Time Thermodynamics of an Autonomous Information Machine},
  author={Chen, Wanyan and Chen, Miao and Ma, Yu-Han},
  journal={arXiv:2604.15953},
  year={2026},
  doi={10.48550/arXiv.2604.15953}
}

@article{qiu20roadmap,
  title={Roadmap on thermodynamics and thermal metamaterials},
  author={Qiu, Yuguang and Nomura, Masahiro and Zhang, Zhongwei and others},
  journal={Front. Phys.},
  volume={20},
  number={6},
  pages={065500},
  year={2026},
  doi={10.15302/frontphys.2025.065500}
}

@misc{xia,
      title={Nonorthogonal-state erasure as the resource behind apparent second-law violations}, 
      author={Xinshu Xia and Hui Hui Qin and Yu-Han Ma and Chang-Pu Sun and Hui Dong},
      year={2026},
      eprint={2608.13881},
      archivePrefix={arXiv},
      primaryClass={quant-ph},
}

\end{document}